\documentclass[a4paper,12pt]{article}

\usepackage{graphicx}
\usepackage{amsmath}
\usepackage{amssymb}
\usepackage{amsfonts}
\usepackage{amsthm}
\usepackage[usenames]{color}
\usepackage[normalem]{ulem} 
\usepackage{cancel} 
\usepackage{cite}
\usepackage{bm}

\renewcommand\sout{\bgroup \color{red} \ULdepth=-.5ex \ULset}

\newcommand{\del}{\partial}

 \makeatletter
 \@addtoreset{equation}{section}
 \makeatother

\newcommand{\group}[1]{\mathrm{#1}}
\newcommand{\e}{\mathrm{e}}
\newcommand{\Transpose}{\mathsf{T}}
\newcommand{\dop}{\mathrm{d}}
\newcommand{\I}{i}
\theoremstyle{definition}

\newtheorem*{proposition*}{Proposition}
\usepackage{hyperref}
\hypersetup{%
  colorlinks=true,%
  citecolor=blue,%
  linkcolor=black,%
  urlcolor=black,%
}
\newcommand{\arXiv}[1]{\href{https://arxiv.org/abs/#1}{arXiv:\textcolor{black}{#1}}}

\allowdisplaybreaks

\date{empty}
\begin{document}
\begin{titlepage}
\null
\begin{flushright}
May, 2026 
\end{flushright}
\vskip 2cm
\begin{center}
{\Large \bf 
Gauge-Dressed Complex Geometry \\[1.5mm]
and T-duality in Heterotic String Theories 
}
\vskip 2cm
\normalsize
\renewcommand\thefootnote{\alph{footnote}}

{\large
Shin Sasaki${}^{1}$\footnote{shin-s(at)kitasato-u.ac.jp}
and
Kenta Shiozawa${}^{2}$\footnote{shiozawa.kenta(at)kitasato-u.ac.jp}
}
\vskip 0.5cm

  {\it
  ${}^{1}$Department of Physics, Kitasato University \\
  Sagamihara 252-0373, Japan
  }

  {\it
  ${}^{2}$Center for Natural Sciences, College of Liberal Arts and
  Sciences, \\ 
  Kitasato University \\
  Sagamihara 252-0373, Japan
  }
\vskip 0.7cm
\begin{abstract}
We study T-duality of $(p,q)$-hermitian geometries in backgrounds
 with non-Abelian gauge fields $A$ in heterotic string theories.
We introduce a gauge-dressed complex geometry characterized
 by a shifted metric $\bar{g} = g + \frac{1}{2} \mathrm{Tr}(A^2)$, 
the closed 2-form $\omega$ and a quasi complex structure satisfying $\bar{J}^2 < 0$, but not necessarily
 $\bar{J}^2 = -1$.
Utilizing the positive and negative chirality half generalized complex-like
 structures constructed by $(\bar{g}, \bar{J})$, 
we derive a heterotic Buscher-like rule for geometric quantities.
We also demonstrate that the gauge-dressed structures can be used to construct an extended Born
 geometry that satisfies algebras of hypercomplex numbers.
\end{abstract}
\end{center}

\end{titlepage}

\newpage
\setcounter{footnote}{0}
\renewcommand\thefootnote{*\arabic{footnote}}
\pagenumbering{arabic}
\tableofcontents

\section{Introduction} \label{sect:introduction}

T-duality is an essential feature for understanding the whole picture of string theory.
Unlike point particle theories,  
T-duality which arises from the fact that the fundamental string has a
finite length scale $l_s \sim \sqrt{\alpha'}$, provides
non-trivial connections among various string theories.
It is well-known that when type II string theories are compactified on the $D$-dimensional torus $T^D$, 
then the T-duality group is given by $\group{O}(D,D)$.
Geometric structures such as the space-time metric $g_{\mu \nu}$, 
the $B$-field $B_{\mu \nu}$, the dilaton $\phi$ and also the RR fields
in supergravity solutions 
are related by the $\group{O}(D,D)$ T-duality transformations.
The explicit transformation formula is known as the Buscher rule~\cite{Buscher:1987sk}.

On the other hand, spacetimes that appear as solutions to supergravity theories 
are known to be target spaces of two-dimensional supersymmetric string sigma models.
It has been shown that the supersymmetry of the two-dimensional
$\mathcal{N} = (2,2)$ sigma model requires complex structures $J$ of the target space.
The target space of $\mathcal{N} = (2,2)$ supersymmetric sigma models 
are generally complex manifolds, which can be either K\"ahler manifolds 
or bi-hermitian manifolds, depending on the presence of the $B$-field \cite{Gates:1984nk}.
Similar results hold for the case of the $\mathcal{N} = (4,4)$ sigma models,
for which the target spaces are hyperk\"ahler or bi-hypercomplex manifolds.
These spaces are characterized by geometric structures, 
such as several almost complex structures $J_{\pm}$ and fundamental 2-forms $\omega_{\pm}$.
Since supersymmetric sigma models that admit isometries are related each other by T-duality, 
the transformation rule of these geometric quantities is also important. 
This issue has mainly been studied in terms of sigma models~\cite{Hassan:1994mq} 
and later in the language of generalized complex geometry~\cite{Gualtieri:2004}.

On the other hand, a manifestly T-duality covariant formalism of the $D$-dimensional type~II supergravities, 
known as double field theory (DFT) has been developed in recent years \cite{Hull:2009mi}.
DFT is defined on a $2D$-dimensional doubled space $\mathcal{M}_{2D}$.
All the dynamical fields and gauge parameters in DFT are defined on $\mathcal{M}_{2D}$ 
and subject to a constraint known as the strong constraint.
Upon imposing the strong constraint, 
a $D$-dimensional physical spacetime $M_D$ is defined in a subspace of $\mathcal{M}_{2D}$.
The spacetime metric $g$, the NSNS $B$-field are packaged into the generalized metric $\mathcal{H}$ in $\mathcal{M}_{2D}$.
It is easily checked that the action of the NSNS sector of type~II supergravities 
are obtained via the imposition of the strong constraint on the DFT action.

In DFT, it is an easy exercise to write down the explicit T-duality
transformation (Buscher rule) for the metric, the $B$-field, the dilaton
$\phi$ and the RR forms.
On the other hand, geometric structures such as the almost complex structures $J_{\pm}$ 
and the fundamental 2-forms $\omega_{\pm}$ 
have been incorporated into $\group{O}(D,D)$ covariant forms in the
context of generalized geometry~\cite{Gualtieri:2004}.
Utilizing the relation between the doubled formalism and the generalized geometry~\cite{Freidel:2017yuv}, 
the explicit T-duality rule for the $(J_{\pm},\omega_{\pm})$ has
been written down for K\"ahler, bi-hermitian, hyperk\"ahler and bi-hypercomplex geometries~\cite{Kimura:2022dma}.
This is the story in type~II theories.

The purpose of this paper is to establish the T-duality rule for the
geometric quantities in heterotic theories.
There are several things we need to be concerned about.
One is that we should incorporate the non-Abelian gauge fields in heterotic supergravities. 
They indeed contribute to the $\alpha'$-corrections in the heterotic
Buscher rule~\cite{Bergshoeff:1995cg, Serone:2005ge, Hohm:2014sxa}.
The other is that we should also take into account the fact that 
the heterotic theories admit unbalanced left and right chiral supersymmetries in their sigma model description.
This is known as the two-dimensional $\mathcal{N} = (p,q)$ supersymmetric sigma models~\cite{Hull:1985jv}.
It has been shown that the target spaces of the $\mathcal{N} = (p,q)$
sigma models are $(p,q)$-hermitian manifolds
admitting $(p-1)$ chiral and $(q-1)$ anti-chiral complex structures $(J_{+,a},J_{-,a'})$ 
($a=1, \ldots, p-1; a' = 1, \ldots, q-1$)~\cite{Howe:1988cj}.
The strong (hyper)K\"ahler with torsion geometry is a specific example of the $(p,q)$-hermitian geometry~\cite{Howe:1996kj}.
The generalized geometry associated with the $(p,q)$-hermitian geometry
has been developed in~\cite{Hull:2018jkr}.
On the other hand,
the DFT formalism is also generalized to heterotic theory in the framework of gauged DFT~\cite{Hohm:2011ex}.
We will utilize these formalisms transversely to reveal the heterotic
T-duality rule for $(p,q)$-hermitian geometry.
In particular, we will discuss the effects of non-Abelian gauge fields on geometries and T-duality.

The organization of this paper is as follows. 
In the next section, we introduce the notion of $(p,q)$-hermitian
geometries and associated $(p,q)$-generalized geometry.
We also discuss the doubled geometry, DFT and their
relation to $(p,q)$-generalized geometry.
In Section \ref{sect:quasi-compl_and_het_TD}, 
we introduce the notion of the gauge-dressed geometry through the 
relation between the $\group{O}(D,D)$ and the $\group{O}(D,D+n)$
formalisms.
Using the doubled formalism, we write down the explicit form of the Buscher-like formula for the complex structures 
and fundamental 2-forms in the $(p,q)$-hermitian geometries with gauge
field backgrounds.
In Section \ref{sec:gauge-dressed_geometry}, we will show that the 
gauge-dressed geometry is utilized to construct Born and generalized 
almost complex structures that constitute the doubled geometry of DFT.
In Section \ref{sec:example}, we show an example of the gauge-dressed geometry.
Section \ref{sec:conclusion} is devoted to the conclusion and discussions.
In Appendix \ref{App:compatible_comp_str}, a brief proof of almost complex structures is shown.

\section{$(p,q)$-hermitian and $(p,q)$-generalized geometries}\label{sect:(p,q)-hermitian}

\subsection{$\mathcal{N} = (p,q)$ sigma models and $(p,q)$-hermitian geometry}
We briefly introduce the $(p,q)$-hermitian geometry relevant to
heterotic sigma models, i.e., the two-dimensional sigma models that have
$\mathcal{N}=(p,q)$ worldsheet supersymmetry.
The target space of such models is known to be a $(p,q)$-hermitian manifold~\cite{Hull:2018jkr}. 
A $(p,q)$-hermitian geometry is defined by the following tuple of data $(M_D, J_{+,a}, J_{-,a'}, g, H)$, 
where $M_D$ is a 
$D$-dimensional smooth manifold, $g$ is a Riemannian metric on $M_D$ and $H$ is a closed 3-form. 
$J_{+,a}$ ($a = 1, \ldots, p-1$) and $J_{-,a'}$ ($a' = 1, \ldots, q-1$) are both complex structures on $M_D$. 
The symbols $+$ and $-$ correspond to the worldsheet chirality.
These complex structures are hermitian with respect to the metric $g$, 
satisfying $g (J_\pm \cdot, J_\pm \cdot) = g(\cdot, \cdot)$, 
where $J_\pm$ collectively denote $J_{+,a}$ and $J_{-,a'}$. 
A triple $(J_{+,a}, J_{-,a'}, g)$ is called a $(p,q)$-hermitian structure. 
Each $J_\pm$ has an associated fundamental 2-form $\omega_\pm = - g J_\pm$. 
Since $\omega_\pm$ is non-degenerate, it is also called an almost symplectic structure. 
Furthermore, these complex structures must be covariantly constant with respect to 
a torsionful Bismut connection $\nabla^{(\pm)} = \nabla^{(0)} \pm \frac{1}{2} g^{-1} H$, i.e., $\nabla^{(\pm)} J_\pm = 0$. 
Here, $\nabla^{(0)}$ is the Levi-Civita connection without the torsion. 
This condition is equivalent to the closure of the supersymmetry algebra. 
Note that when $p = 4$, $J_{+,a}$ satisfies the quaternion relation $J_{+,1} J_{+,2} = J_{+,3}$. 
The same holds for $q = 4$ and $J_{-,a'}$.

There are important examples for the $(p,q)$-hermitian geometries in heterotic and type~II string theories.
For instance, the target space geometry of
$\mathcal{N} = (2,1)$ or $\mathcal{N} = (2,0)$ models
naturally arises in the supersymmetric backgrounds of heterotic compactifications with non-vanishing $H$-flux, 
known as the Strominger system~\cite{Strominger:1986uh, Hull:1986kz}.
In such backgrounds, the geometry is characterized by a hermitian metric with totally antisymmetric torsion, 
closely related to the strong K\"ahler with torsion (SKT) geometries~\cite{Hull:1985zy}. 
Furthermore, the $(2,2)$-hermitian geometry is equivalent to the bi-hermitian geometry~\cite{Gates:1984nk, Howe:1984fak}. 
The $(4,4)$-hermitian geometry corresponds to the bi-hypercomplex geometry, 
which typically appears in the background of the type~II NS5-branes.

\subsection{$(p,q)$-generalized geometry}\label{sect:(p,q)_gen_geom}
A concept related to $(p,q)$-hermitian geometry is $(p,q)$-generalized complex geometry~\cite{Hull:2018jkr}. 
Generalized complex geometry is defined on $\mathbb{T}M_D = TM_D \oplus
T^*M_D$, called the generalized tangent bundle~\cite{Gualtieri:2004}. 
Here, $M_D$ is a $D$-dimensional manifold, and $TM_D$ and $T^*M_D$ are the tangent bundle and cotangent bundle of $M_D$, respectively.
A natural symmetric bilinear form $\eta (X+\alpha, Y+\beta) = \beta(X) + \alpha(Y)$ is defined on $\mathbb{T}M_D$, 
where $X, Y \in \Gamma (TM_D)$ and $\alpha, \beta \in \Gamma (T^*M_D)$. 
$\eta$ is also called an $\group{O}(D,D)$-invariant metric.
Generalized geometry gives a natural framework for T-duality by
unifying the metric $g$ and the $B$-field on the $D$-dimensional manifold
$M_D$ into a $2D \times 2D$ generalized metric on $\mathbb{T}M_D$:
\begin{align}
\mathcal{H} = 
\begin{pmatrix}
g - B g^{-1} B & B g^{-1} \\
- g^{-1} B & g^{-1}
\end{pmatrix}
\end{align}
on which T-duality transformations are
manifested as $\mathrm{O}(D,D)$ rotations. 
This formalism is closely related to the $2D$-dimensional doubled geometry $\mathcal{M}_{2D}$.
Indeed, the generalized tangent bundle $\mathbb{T}M_D$ and the doubled tangent bundle
$T\mathcal{M}_{2D}$ are identified under the
strong constraint of type~II double field theory \cite{Freidel:2018tkj, Freidel:2013zga, Rudolph:2019fir, Kimura:2022dma,
Kimura:2022jyp}.
We will discuss the heterotic case in the next section.

An endomorphism $\mathcal{J} \in \operatorname{End}(\mathbb{T}M_D)$ of the generalized tangent bundle 
that satisfies $\mathcal{J}^2 = -1$ and preserves $\eta$, i.e., $\eta (\mathcal{J} \cdot, \mathcal{J} \cdot) = \eta (\cdot, \cdot)$, 
is called a generalized almost complex structure~\cite{Gualtieri:2004}.
Employing the 3-form $H = {\dop}B$ associated with the 2-form $B$, the algebraic structure 
on $\mathbb{T}M_D$ is governed by the $H$-twisted Courant bracket~\cite{Severa:2001qm}.
The integrability of a generalized almost complex structure 
$\mathcal{J}$ is defined by the involutivity
of its eigenbundles under the $H$-twisted Courant bracket~\cite{Gualtieri:2004}. 
The integrable generalized almost complex structure $\mathcal{J}$ is called a generalized complex structure.

A pair of commutative generalized complex structures $(\mathcal{J}_1, \mathcal{J}_2)$
is called a generalized K\"ahler structure~\cite{Gualtieri:2010fd}. 
The product $\mathcal{G} = - \mathcal{J}_1 \mathcal{J}_2$ is called a chiral structure and satisfies $\mathcal{G}^2 = 1$.
The projection operator onto the eigenbundle corresponding to the eigenvalues $\pm 1$ of the chiral structure $\mathcal{G}$
is defined as $P_\pm = \frac{1}{2} (1 \pm \mathcal{G})$, which is called the chiral projection operator. 
The chiral projection operators split the generalized tangent bundle into the $\pm 1$-eigenbundles $C_\pm$, i.e., $\mathbb{T}M_D = C_+ \oplus C_-$. 

We next introduce a half generalized complex structure.
This is defined as a generalized almost complex structure $\mathcal{I}^{(\pm)}$ 
on the chiral subbundles $C_\pm \subset \mathbb{T}M_D$ satisfying $(\mathcal{I}^{(\pm)})^2 C_\pm = - C_\pm$ 
and $\mathcal{I}^{(\pm)} C_\mp = 0$~\cite{Hull:2018jkr}. 
Note that $\mathcal{I}^{(\pm)}$ is not a generalized complex structure on $\mathbb{T}M_D$, 
since $(\mathcal{I}^{(\pm)})^2 = -1$ does not hold on the whole space of $\mathbb{T}M_D$. 
$\mathcal{I}^{(+)}$ ($\mathcal{I}^{(-)}$) is called a positive (negative)
chirality half generalized almost complex structure.
Since $(\mathcal{I}^{(+)})^2 = -1$ on $C_{+}$, 
this is decomposed as $C_+ = L_+ \oplus \bar{L}_+$, 
where $L_+$ and $\bar{L}_+$ are the $+\I$- and $-\I$-eigenbundles of $\mathcal{I}^{(+)}$, respectively. 
If $L_+$ is involutive under the $H$-twisted Courant bracket, 
then $\mathcal{I}^{(+)}$ is called a positive chirality half generalized complex structure. 
A similar construction exists for negative chirality. 

A pair $(\mathcal{I}^{(+)}_a, \mathcal{I}^{(-)}_{a'})$ consisting of 
$(p-1)$ positive chirality half generalized complex structures $\mathcal{I}^{(+)}_a$ 
and $(q-1)$ negative chirality half generalized complex structures $\mathcal{I}^{(-)}_{a'}$ 
is called a $(p,q)$-generalized complex structure.
$\mathcal{I}^{(+)}_a$ satisfies the quaternion algebra when $p = 4$, 
and the same is true
for $\mathcal{I}^{(-)}_{a'}$ when $q = 4$. 
A $(p,q)$-generalized complex geometry is defined by the following tuple of data
$(\mathbb{T}M_D, \mathcal{I}^{(+)}_a, \mathcal{I}^{(-)}_{a'}, \mathcal{G}, H)$,
given by the generalized tangent bundle $\mathbb{T}M_D$ equipped with 
the $(p,q)$-generalized complex structure and a chiral structure, 
and a 3-form $H$ on $M_D$.
It is known that the $(p,q)$-generalized complex geometry and the $(p,q)$-hermitian geometry are equivalent~\cite{Hull:2018jkr}. 

If $p = q$, generalized (almost) complex structures on $\mathbb{T}M_D$ are defined as 
\begin{align}
\mathcal{J}^{(\pm)}_a &= \frac{1}{2} \left( \mathcal{I}^{(+)}_a \pm
 \mathcal{I}^{(-)}_a \right)
\quad
(a = 1, \ldots, p-1).
\end{align}
When $(p,q) = (2,2)$ and $\mathcal{J}^{(\pm)}$ is integrable, 
the pair $(\mathcal{J}^{(+)}, \mathcal{J}^{(-)})$ 
defines a generalized K\"ahler structure. 
Furthermore, when $(p,q) = (4,4)$, the structure $\mathcal{J}^{(\pm)}_a$ has the following relation:
\begin{alignat}{2}
\mathcal{J}^{(+)}_{a} \mathcal{J}^{(+)}_{b} 
&= - \delta_{ab} \mathbf{1}_{2D} + \epsilon_{abc} \mathcal{J}^{(+)}_{c}, &\qquad 
\mathcal{J}^{(+)}_{a} \mathcal{J}^{(-)}_{b} 
&= \delta_{ab} \mathcal{G} + \epsilon_{abc} \mathcal{J}^{(-)}_{c}, 
\notag \\
\mathcal{J}^{(-)}_{a} \mathcal{J}^{(-)}_{b} 
&= - \delta_{ab} \mathbf{1}_{2D} + \epsilon_{abc} \mathcal{J}^{(+)}_{c}, &\qquad 
\mathcal{J}^{(-)}_{a} \mathcal{J}^{(+)}_{b} 
&= \delta_{ab} \mathcal{G} + \epsilon_{abc} \mathcal{J}^{(-)}_{c}.
\label{eq:generalized_hyperkahler_algebra}
\end{alignat}
If all $\mathcal{J}^{(\pm)}_a$ are integrable, 
then the pair of structures $(\mathcal{J}^{(+)}_a, \mathcal{J}^{(-)}_a)$ is 
the generalized hyperk\"ahler structure~\cite{Bredthauer:2006sz}.
The algebra \eqref{eq:generalized_hyperkahler_algebra} is known as that
of a hypercomplex number, the split-bi-quaternions \cite{Kimura:2022jyp}.

We stress that the defining properties of these generalized quantities
come from the algebraic relation $\omega = - g J$, or equivalently $J =
- g^{-1} \omega$.
In heterotic string theories, however, the presence of non-Abelian gauge backgrounds 
modifies the natural metric appearing in the T-duality covariant formulation. 
This motivates the gauge-shifted reformulation discussed in the next section.

\section{Gauge-dressed geometry and heterotic T-duality}\label{sect:quasi-compl_and_het_TD}

In the previous section, we had not considered the gauge sector in
heterotic theories. 
Indeed, it is known that non-trivial gauge fields modify the T-duality
rules for heterotic theories~\cite{Bergshoeff:1995cg, Serone:2005ge} and
there should be modifications to the generalized geometry.
In this section, we discuss, through the doubled geometry, 
how the gauge sector modifies the generalized complex geometry.

\subsection{$\group{O}(D,D+n)$ vs. $\group{O}(D,D)$}
In the low-energy effective supergravity of heterotic string theories, 
the massless spectrum contains the metric $g$, the Kalb--Ramond field $B$, the dilaton $\phi$, 
and the non-Abelian gauge fields $A$ belonging to the gauge group $G$, 
which is either $\group{E}_8 \times \group{E}_8$ or $\group{SO}(32)$ (where $\dim G = 496$). 
When the theory is compactified on a $D$-dimensional torus $T^D$, 
the continuous symmetry of the resulting effective action 
and the corresponding T-duality depend strongly on the gauge field backgrounds. 
If one turns on generic Wilson lines along the $D$-torus directions, 
the gauge group $G$ is generically broken down to 
its maximal Cartan subgroup $\group{U}(1)^{16}$, and the Narain lattice construction exhibits 
the well-known $\group{O}(D,D+16)$ T-duality symmetry~\cite{Narain:1986am, Narain:1985jj}.
On the other hand, when there are no Wilson lines and the full gauge
group $G$ is retained, the T-duality symmetry is given by
$\group{O}(D,D)$~\cite{Hohm:2014sxa}.
In order to realize the manifest $\group{O}(D,D)$ symmetry in heterotic
supergravity, the so-called gauged DFT has been
studied~\cite{Garcia-Fernandez:2013gja, Hohm:2011ex}.
The fields in gauged DFT consist of the $\group{O} (D,D+n)$ generalized
metric $\widehat{\mathcal{H}}$ and the generalized
dilation~\cite{Hohm:2011ex}.
Note that $n=496$ for $G = \group{E}_8 \times \group{E}_8$ or $\group{SO}(32)$.
The generalized metric is parametrized as
\begin{align}
\widehat{\mathcal{H}} 
&= 
\begin{pmatrix}
g + C^{\Transpose} g^{-1} C + A \kappa A^{\Transpose} & - C^{\Transpose} g^{-1} & C^{\Transpose} g^{-1} A + A \\
- g^{-1} C & g^{-1} & - g^{-1} A \\
A^{\Transpose} g^{-1} C + A^{\Transpose} & - A^{\Transpose} g^{-1} & \kappa^{-1} + A^{\Transpose} g^{-1} A
\end{pmatrix},
\label{eq:O(D,D+n)_gen_met}
\end{align}
where $C = B + \frac{1}{2} A\kappa A^{\Transpose}$.\footnote{%
Here, the gauge sector appears as $\alpha'$-corrections. 
We have rescaled the gauge field $A \mapsto \frac{1}{\sqrt{\alpha'}} A$ and 
we do not show $\alpha'$ explicitly in the following. 
} 
Here, $\kappa = (\kappa_{\alpha\beta})$ is the Cartan--Killing metric of the heterotic gauge group with $\alpha = 1, \ldots, n$, 
and $\bullet^{\Transpose}$ denotes the transpose of a matrix. 
The generalized metric includes the metric $g$, the $B$-field, and the gauge field $A$; 
consequently, the $\group{O}(D, D+n)$ framework provides a unified description of the gravity and gauge sectors.
After introducing the global $\group{O} (D,D+n)$ symmetry, a part of it
is gauged to give the gauge group $G$ and $\group{O} (D,D+n)$
is broken down to its subgroup $\group{O}
(D,D)$~\cite{Garcia-Fernandez:2013gja, Hohm:2011ex}.
In order to rewrite everything in $\group{O}(D,D)$, we need to reorganize the fields. 
The gravity and gauge sectors are encoded in 
the generalized metric $\overline{\mathcal{H}}{}^B$
and the $\group{O}(D,D)$ vector $\mathcal{C}$~\cite{Hohm:2014sxa}:
\begin{align}
\overline{\mathcal{H}}{}^B 
&= 
\begin{pmatrix}
\bar{g} - B \bar{g}^{-1} B & B \bar{g}^{-1} \\
- \bar{g}^{-1} B & \bar{g}^{-1} 
\end{pmatrix}, 
\qquad
\mathcal{C} = \frac{1}{2}
\begin{pmatrix}
- B \bar{g}^{-1} A^{\Transpose} \kappa + A^{\Transpose} \kappa \\
- \bar{g}^{-1} A^{\Transpose} \kappa 
\end{pmatrix}, 
\label{eq:mod_gen_met_and_O(D,D)_vector}
\end{align}
where $\bar{g}$ is the modified metric~\cite{Maharana:1992my} given by 
\begin{align}
\bar{g} &= g + \frac{1}{2} A \kappa A^{\Transpose},
\label{eq:shift_metric}
\end{align}
which we call the \textit{gauge-shifted} spacetime metric. 
The overline notation $\overline{\mathcal{H}}{}^B$ explicitly indicates that $\bar{g}$ in the parametrization~\eqref{eq:mod_gen_met_and_O(D,D)_vector} is the gauge-shifted spacetime metric.
We refer to $\overline{\mathcal{H}}{}^B$ as a gauge-shifted generalized metric. 
The gauge-shifted generalized metric satisfies the $\group{O}(D,D)$ constraint 
$\overline{\mathcal{H}}{}^B \eta^{-1} \overline{\mathcal{H}}{}^B = \eta$, 
and the $\group{O}(D,D)$ vector $\mathcal{C}$ is constrained by 
$(1 + \overline{\mathcal{H}}{}^B \eta^{-1}) \mathcal{C} = 0$.
It is shown in~\cite{Hohm:2014sxa} that the $\group{O}(D,D)$ formulation by
$(\overline{\mathcal{H}}{}^B, \mathcal{C})$
is equivalent to the $\group{O}(D,D+n)$ formulation by
$\widehat{\mathcal{H}}$. 
In conventional DFT, the action is fully described by the generalized metric and the generalized dilaton. 
However, this is no longer sufficient in the $\group{O}(D,D)$-covariant
formulation of heterotic DFT, where an additional $\group{O}(D,D)$
vector $\mathcal{C}$ must be introduced.

The $\group{O}(D,D)$-covariant heterotic DFT is defined on the $2D$-dimensional space $\mathcal{M}_{2D}$. 
The gauge-shifted generalized metric $\overline{\mathcal{H}}{}^B$ and
the $\group{O}(D,D)$ vector $\mathcal{C}$ are regarded as fields on the
doubled tangent bundle $T\mathcal{M}_{2D}$. 

The T-duality relation of the 
fields 
$\bar{g}$, $B$ and $\phi$
is well-captured by an $\group{O}(D,D)$ rotation of the quantities~\eqref{eq:mod_gen_met_and_O(D,D)_vector}. 
By applying an $\group{O}(D,D)$ factorized T-duality transformation
along the isometry direction $y$, 
one reproduces the heterotic Buscher rule~\cite{Bergshoeff:1995cg, Serone:2005ge}: 
\begin{alignat}{3}
\bar{g}'_{ij} &= \bar{g}_{ij} - \frac{\bar{g}_{iy} \bar{g}_{jy} - B_{iy} B_{jy}}{\bar{g}_{yy}}, & \qquad 
\bar{g}'_{iy} &= \frac{B_{iy}}{\bar{g}_{yy}}, & \qquad
\bar{g}'_{yy} &= \frac{1}{\bar{g}_{yy}}, 
\notag \\
B'_{ij} &= B_{ij} - \frac{B_{iy} \bar{g}_{jy} - \bar{g}_{iy} B_{jy}}{\bar{g}_{yy}}, & \qquad 
B'_{iy} &= \frac{\bar{g}_{iy}}{\bar{g}_{yy}} &\qquad
\e^{2\phi'} &= \frac{\e^{2\phi}}{\bar{g}_{yy}},
\label{eq:hetero_Buscher_rule_1}
\end{alignat}
where $i,j \not= y$. 
The explicit derivation of the equation~\eqref{eq:hetero_Buscher_rule_1} is given in Section~\ref{sect:T-duality_rules}. 
The heterotic Buscher rule takes the same form as the type~II Buscher
rule with the metric $g$ replaced by the gauge-shifted metric $\bar{g}$.
The rule~\eqref{eq:hetero_Buscher_rule_1} suggests that the geometry is
effectively governed by the gauge-shifted metric $\bar{g}$ modified by the
gauge field background in heterotic theories.
This motivates us to introduce the notion of {\it gauge-dressed geometry} in the
next section.

\subsection{Quasi complex structures and gauge-dressed geometry}
In order to find the heterotic T-duality transformation of (almost)
complex structures, we develop a geometric framework that is compatible with the gauge-shifted
metric $\bar{g}$.

As a first step, we define a new geometric object by combining the
gauge-shifted metric $\bar{g}$ and the fundamental 2-form $\omega$.
We assume that $\bar{g}$ is positive-definite and
hence non-degenerate.\footnote{
Throughout this paper, we assume that the spacetime metric $g$ and the
Cartan metric $\kappa$ have positive-definite Euclidean signatures, which
implies that $\bar{g}$ is positive-definite and hence non-degenerate.}
We introduce a map $\bar{J} : TM_D \to TM_D$, which we call the \textit{quasi complex structure}, 
defined by 
\begin{align}
\bar{J} &= - \bar{g}^{-1} \omega.
\label{eq:defn_quasi_complex_strc}
\end{align}
In this definition, the gauge field contribution is incorporated into
$\bar{J}$ through the gauge-shifted metric $\bar{g}$, while the
fundamental 2-form $\omega$ remains unchanged. 
In particular, the latter is given by the original metric $g$ and a complex structure $J$, namely $\omega = - g J$.
This construction is motivated by the fact that the compatibility
between $\bar{g}$ and $\omega$ is the essential information for encoding
the $(p,q)$-hermitian structure into an $\group{O}(D,D)$-covariant
generalized geometric structure.\footnote{An alternative approach is to
incorporate the gauge field contribution into the fundamental 2-form,
defining $\bar{\omega} = - \bar{g} J$. However, in this case,
$\bar{\omega}$ is generally not antisymmetric. This is not a desirable
property for encoding the $(p,q)$-hermitian structure into generalized
structures.}
We can show that $\bar{J}^2$ is negative-definite in general
(see Appendix \ref{App:compatible_comp_str}), but 
it is important to notice that $\bar{J}$ generally does not satisfy the
algebraic property of an ordinary almost complex structure; that is,
$\bar{J}^2 \neq -1$ as long as the gauge field background is
non-trivial.
For this reason, we refer to $\bar{J}$ as a \textit{quasi} complex structure. 
Although $\bar{J}$ loses the standard property $J^2 = -1$, it serves as
an essential building block to construct the $\group{O}(D,D)$-covariant
generalized structures in the following discussion.

As a next step, we embed the quasi complex structure $\bar{J}$ into $\mathbb{T}M_D$. 
Motivated by the embedding of an almost complex structure into a
generalized almost complex structure (the so-called Gualtieri
map~\cite{Gualtieri:2004}), we define the \textit{generalized quasi complex structure}: 
\begin{align}
\overline{\mathcal{J}}_{\bar{J}} 
&= 
\begin{pmatrix}
\bar{J} & 0 \\
0 & - \bar{J}^{\Transpose}
\end{pmatrix}
= 
\begin{pmatrix}
- \bar{g}^{-1} \omega & 0 \\
0 & - \omega \bar{g}^{-1}
\end{pmatrix}.
\end{align}
Here, for the convenience of later calculations, we have expressed the generalized quasi complex structure in terms of $\bar{g}$ and $\omega$, using Equation~\eqref{eq:defn_quasi_complex_strc}. 
Note that $(\overline{\mathcal{J}}_{\bar{J}})^2 \neq -1$ in general.

As a final step, we discuss the generalized structures relevant to $(p,q)$-hermitian geometry with gauge fields. 
An important point is that, corresponding to a $(p,q)$-hermitian structure $(g, J_{+,a}, J_{-,a'})$ on the target space, there exist the gauge-shifted metric $\bar{g}$, $(p-1)$ quasi complex structures $\bar{J}_{+,a}$, and $(q-1)$ quasi complex structures $\bar{J}_{-,a'}$. 
For each of these quasi complex structures, there also exists a corresponding generalized quasi complex structure.
We first define a modified chiral structure using the gauge-shifted metric $\bar{g}$ as in
\begin{align}
\overline{\mathcal{G}} 
&= 
\begin{pmatrix}
0 & \bar{g}^{-1} \\
\bar{g} & 0
\end{pmatrix}. 
\end{align}
Since $\overline{\mathcal{G}}{}^2 = 1$, we define the modified chiral projection operators as 
\begin{align}
\overline{P}_\pm 
&= \frac{1}{2} \big( 1 \pm \overline{\mathcal{G}} \big).
\end{align} 
By acting these projection operators on the generalized quasi complex structures 
$\overline{\mathcal{J}}_{\bar{J}_{+,a}}$ and $\overline{\mathcal{J}}_{\bar{J}_{-,a'}}$, 
we define the half generalized quasi complex structures\footnote{%
Note that the notation $\pm$ in $\overline{\mathcal{I}}^{(\pm)}$ corresponds to the spacetime chirality defined by $\overline{P}_\pm$ while those in $\bar{J}_\pm$ to the worldsheet chirality. 
We employ the convention in~\eqref{eq:mod_half_gen_quasi_compl} but usually these chiralities are chosen to be the same in the literature. This is a matter of convention. 
} as 
\begin{align}
\overline{\mathcal{I}}^{(-)}_a 
&= \overline{P}_- \big( \overline{\mathcal{J}}_{\bar{J}_{+,a}} \big) \overline{P}_-, \qquad 
\overline{\mathcal{I}}^{(+)}_{a'} 
= \overline{P}_+ \big( \overline{\mathcal{J}}_{\bar{J}_{-,a'}} \big) \overline{P}_+. 
\label{eq:mod_half_gen_quasi_compl}
\end{align}
The explicit matrix representations of 
$\overline{\mathcal{I}}^{(-)}_a$ and $\overline{\mathcal{I}}^{(+)}_{a'}$ are obtained as 
\begin{align}
\overline{\mathcal{I}}^{(-)}_a 
&= \frac{1}{2} 
\begin{pmatrix}
- \bar{g}^{-1} \omega_{+,a} & \bar{g}^{-1} \omega_{+,a} \bar{g}^{-1} \\
\omega_{+,a} & - \omega_{+,a} \bar{g}^{-1}
\end{pmatrix},  
\qquad
\overline{\mathcal{I}}^{(+)}_{a'} 
= - \frac{1}{2} 
\begin{pmatrix}
\bar{g}^{-1} \omega_{-,a'} & \bar{g}^{-1} \omega_{-,a'} \bar{g}^{-1} \\
\omega_{-,a'} & \omega_{-,a'} \bar{g}^{-1}
\end{pmatrix}.
\label{eq:my_note_33}
\end{align}
Therefore, half generalized quasi complex structures $\overline{\mathcal{I}}^{(\pm)}$ can be expressed by the gauge-shifted metric $\bar{g}$ 
and the fundamental 2-form $\omega_\pm$, just like $\overline{\mathcal{J}}_{\bar{J}}$. 

If a $B$-field exists, the generalized structure must be twisted by the
$B$-transformation~\cite{Gualtieri:2004}, 
\begin{align}
\e^B 
&= 
\begin{pmatrix}
1 & 0 \\
-B & 1
\end{pmatrix},
\label{eq:B-transf_mat}
\end{align}
which is an $\group{O}(D,D)$ matrix. 
The modified chiral structure is twisted by $\overline{\mathcal{G}}{}^B = \e^{-B} \overline{\mathcal{G}} \e^B$. 
Since the modified chiral projection operators behave similarly, the half generalized quasi complex structures are also twisted as 
\begin{align}
\overline{\mathcal{I}}{}^{B(-)}_a 
&= \e^{-B} \overline{\mathcal{I}}^{(-)}_a \e^B, \qquad 
\overline{\mathcal{I}}{}^{B(+)}_{a'} 
= \e^{-B} \overline{\mathcal{I}}^{(+)}_{a'} \e^B.
\label{eq:half_gen_compl_like_strc_with_B}
\end{align}
A key property of $\overline{\mathcal{I}}{}^{B(\pm)}$ is that the
block-matrix structure of their components, expressed in terms of
$\bar{g}$, $\omega$, and $B$, is preserved under T-duality
transformations.
Since the relevant geometric quantities are modified by the gauge
fields, we call this framework {\it the gauge-dressed (generalized) geometry}.
In the next subsection, we derive the T-duality transformation of the
complex structures $J_{+,a}$, $J_{-,a'}$ in gauge field backgrounds,
using the quantities introduced in~\eqref{eq:half_gen_compl_like_strc_with_B}.

\subsection{T-duality rules for $(p,q)$-hermitian geometry}\label{sect:T-duality_rules}
We discuss the T-duality transformation rules for $(p,q)$-hermitian geometry with gauge fields. 
Due to the presence of gauge fields, the Buscher rule receives modifications from them. 
We first re-derive the heterotic Buscher
rule~\eqref{eq:hetero_Buscher_rule_1}
in the gauge-dressed generalized geometry.
By using the gauge-shifted generalized metric~\eqref{eq:mod_gen_met_and_O(D,D)_vector}
and the factorized T-duality matrix, 
\begin{align}
h_y &= 
\begin{pmatrix}
1 - t_y & t_y \\
t_y & 1 - t_y
\end{pmatrix} 
\in \group{O}(D,D), \qquad 
(t_y)_{\mu\nu} = \delta_{\mu y} \delta_{y\nu}, \qquad 
\mu = (i, y),
\label{eq:factorized_T-duality_matrix}
\end{align}
the T-duality transformation of the gauge-shifted generalized metric is given as 
\begin{align}
\overline{\mathcal{H}}{}^B
\
\longmapsto
\
(\overline{\mathcal{H}}{}^B)' 
&= (h_y)^{\mathsf{T}} \, \overline{\mathcal{H}}{}^B \, h_y,
\label{eq:T-dual_gen_met_1}
\end{align}
where $(\overline{\mathcal{H}}{}^B)' = \overline{\mathcal{H}}{}^B (\bar{g}', B')$ is the gauge-shifted generalized metric parametrized 
by $\bar{g}'$ and $B'$. 
By solving the equation~\eqref{eq:T-dual_gen_met_1} for $\bar{g}'$ and $B'$, 
we obtain the equation~\eqref{eq:hetero_Buscher_rule_1}.
We next derive the heterotic Buscher rule for the gauge field. 
In the same way, the $\mathcal{C}$-vector \eqref{eq:mod_gen_met_and_O(D,D)_vector} is transformed by 
the factorized T-duality matrix to $\mathcal{C}' = (h_y)^{\mathsf{T}} \, \mathcal{C}$,
and from this we obtain 
\begin{align}
A'_i{}^\alpha &= A_i{}^\alpha - A_y{}^\alpha \frac{\bar{g}_{iy} + B_{iy}}{\bar{g}_{yy}}, \qquad 
A'_y{}^\alpha = - \frac{A_y{}^\alpha}{\bar{g}_{yy}},
\label{eq:hetero_Buscher_rule_2}
\end{align}
where $\alpha$ is the index for the adjoint representation of the gauge
group $G$.
By using the T-duality transformation rules for the gauge-shifted metric and the gauge fields, 
we obtain the heterotic T-duality transformation rule for the original metric $g_{\mu\nu}$, 
\begin{align}
g'_{ij} &= g_{ij} - g_{iy} \frac{\bar{E}_{jy}}{\bar{E}_{yy}} - g_{jy} \frac{\bar{E}_{iy}}{\bar{E}_{yy}} + g_{yy} \frac{\bar{E}_{iy} \bar{E}_{jy}}{\bar{E}_{yy}^2},
\qquad
g'_{iy} = g_{yy} \frac{\bar{E}_{iy}}{\bar{E}_{yy}^2} - \frac{g_{iy}}{\bar{E}_{yy}}, 
\qquad
g'_{yy} = \frac{g_{yy}}{\bar{E}_{yy}^2},
\end{align}
where $\bar{E}_{\mu\nu} = \bar{g}_{\mu\nu} + B_{\mu\nu}$. 
As a corollary to the derivation of the formula~\eqref{eq:hetero_Buscher_rule_1}, the T-duality transformation rule for the inverse metric is also obtained as
\begin{align}
(g')^{ij} 
&= g^{ij}, \qquad 
(g')^{iy} 
= - g^{i\rho} \bar{E}_{\rho y}, \qquad 
(g')^{yy} 
= g^{\rho\sigma} \bar{E}_{\rho y} \bar{E}_{\sigma y},
\label{eq:T-dual_inv_metric}
\end{align}
where $\rho = (k, y)$ and $\sigma = (l, y)$. 

We then consider the T-duality transformation rules for $(J_\pm, \omega_\pm)$ in $(p,q)$-hermitian geometry with gauge fields. 
Using the factorized T-duality matrix~\eqref{eq:factorized_T-duality_matrix}, 
the gauge-dressed 
positive and negative chirality half generalized quasi complex structures~\eqref{eq:half_gen_compl_like_strc_with_B} are transformed as 
\begin{align}
(\overline{\mathcal{I}}{}^{B(-)}_{a})'
&= \overline{\mathcal{I}}{}^{B(-)}_{a} (\omega'_{+,a}, \bar{g}', B') 
= h_y \, \overline{\mathcal{I}}{}^{B(-)}_{a} \, h_y, 
\notag \\ 
(\overline{\mathcal{I}}{}^{B(+)}_{a'})'
&= \overline{\mathcal{I}}{}^{B(+)}_{a'} (\omega'_{-,a'}, \bar{g}', B') 
= h_y \, \overline{\mathcal{I}}{}^{B(+)}_{a'} \, h_y. 
\label{eq:posi_nega_chiral_T-dual_before_after_relation}
\end{align}
Note that the positive and negative chirality sectors are calculated independently. 
By calculating each right-hand side of the equation~\eqref{eq:posi_nega_chiral_T-dual_before_after_relation} 
and applying the Buscher rule~\eqref{eq:hetero_Buscher_rule_1}, 
we obtain the T-duality transformation rule for 
the fundamental 2-forms $\omega_\pm$: 
\begin{alignat}{2}
(\omega'_{+,a})_{ij} 
&= (\omega_{+,a})_{ij} - \frac{(\omega_{+,a})_{iy} \bar{E}^+_{jy} - \bar{E}^+_{iy} (\omega_{+,a})_{jy}}{\bar{E}_{yy}}, &\qquad 
(\omega'_{+,a})_{iy}
&= - \frac{(\omega_{+,a})_{iy}}{\bar{E}_{yy}}, 
\notag \\
(\omega'_{-,a'})_{ij} 
&= (\omega_{-,a'})_{ij} - \frac{(\omega_{-,a'})_{iy} \bar{E}^-_{jy} - \bar{E}^-_{iy} (\omega_{-,a'})_{jy}}{\bar{E}_{yy}}, &\qquad 
(\omega'_{-,a'})_{iy}
&= + \frac{(\omega_{-,a'})_{iy}}{\bar{E}_{yy}}, 
\label{eq:hetero_Buscher-like_for_omega}
\end{alignat}
where $\bar{E}^\pm_{\mu\nu} = \bar{g}_{\mu\nu} \pm B_{\mu\nu}$, 
and we denoted $\bar{E}_{yy} = \bar{E}_{yy}^\pm$. 
By using the formulas~\eqref{eq:T-dual_inv_metric} and~\eqref{eq:hetero_Buscher-like_for_omega}, and the fact that $J_\pm = - g^{-1} \omega_\pm$, 
we find the T-duality transformation rule for 
the ordinary almost complex structures $J_\pm$:
\begin{alignat}{2}
(J'_{+,a})^i{}_j &= (J_{+,a})^i{}_j - \frac{(J_{+,a})^i{}_y \bar{E}^+_{jy}}{\bar{E}_{yy}}, &\qquad 
(J'_{+,a})^i{}_y &= - \frac{(J_{+,a})^i{}_y}{\bar{E}_{yy}},
\notag \\
(J'_{+,a})^y{}_j &= - \bar{E}^+_{\rho y} \left( (J_{+,a})^\rho{}_j - (J_{+,a})^\rho{}_y \frac{\bar{E}^+_{jy}}{\bar{E}_{yy}} \right), &\qquad 
(J'_{+,a})^y{}_y &= \frac{(J_{+,a})^\rho{}_y \bar{E}^+_{\rho y}}{\bar{E}_{yy}}, 
\notag \\
(J'_{-,a'})^i{}_j &= (J_{-,a'})^i{}_j - \frac{(J_{-,a'})^i{}_y \bar{E}^-_{jy}}{\bar{E}_{yy}}, &\qquad 
(J'_{-,a'})^i{}_y &= + \frac{(J_{-,a'})^i{}_y}{\bar{E}_{yy}},
\notag \\
(J'_{-,a'})^y{}_j &= + \bar{E}^-_{\rho y} \left( (J_{-,a'})^\rho{}_j - (J_{-,a'})^\rho{}_y \frac{\bar{E}^-_{jy}}{\bar{E}_{yy}} \right), &\qquad 
(J'_{-,a'})^y{}_y &= \frac{(J_{-,a'})^\rho{}_y \bar{E}^-_{\rho y}}{\bar{E}_{yy}}, 
\label{eq:hetero_Buscher-like_for_J}
\end{alignat}
where $\rho = (k, y)$.
We emphasize that, in deriving the equations~\eqref{eq:hetero_Buscher-like_for_omega} and~\eqref{eq:hetero_Buscher-like_for_J}, no conditions required other than that $\omega_\pm$ is an antisymmetric tensor, $\bar{g}$ is a symmetric tensor, and $\bar{g}^{-1}$ exists. 
The equations~\eqref{eq:hetero_Buscher-like_for_J} and~\eqref{eq:hetero_Buscher-like_for_omega} resemble 
the Buscher rules~\eqref{eq:hetero_Buscher_rule_1} and~\eqref{eq:hetero_Buscher_rule_2} for $(g, B, A)$ in heterotic theories, hence we call them the heterotic Buscher-like rules for $(J_\pm, \omega_\pm)$. 
The approach here is based on the doubled formalism studied in type~II
theory~\cite{Kimura:2022dma} but here 
we developed a refined method by the gauge-dressed geometry.
We stress that the result \eqref{eq:hetero_Buscher-like_for_J} correctly reproduces the
one previously obtained by sigma model analysis~\cite{Hassan:1994mq}.

\section{Algebra of gauge-dressed geometry in doubled space} \label{sec:gauge-dressed_geometry}
In this section, we investigate gauge-dressed geometry from the
viewpoint of the T-duality covariant formalism---the doubled (or DFT) formalism
which possesses $\group{O}(D,D)$ as a manifest symmetry.
This is defined on
the $2D$-dimensional doubled space $\mathcal{M}_{2D}$ endowed with a Born structure
\cite{Freidel:2017yuv, Freidel:2018tkj}.
In order to specify a $D$-dimensional physical spacetime $M_D$ in the $2D$-dimensional
doubled manifold $\mathcal{M}_{2D}$, we define a para-hermitian structure
$\mathcal{K}$ whose square is the identity on $T \mathcal{M}_{2D}$.
When a $\mathcal{K} = 1$ eigenbundle $L$ and a $\mathcal{K} = -1$
eigenbundle $\tilde{L}$ are integrable, the doubled coordinate is
decomposed as $\mathbb{X}^{M} = (x^{\mu}, \tilde{x}_{\mu})$
and a subspace defined by $\tilde{x} =
\text{const}.$ is identified as a physical spacetime $M_D$.
A generalized metric $\mathcal{H}$ together with the $\group{O}(D,D)$-invariant metric $\eta$ defines a product known as the chiral
structure $\mathcal{G} = \eta^{-1} \mathcal{H}$ satisfying
$\mathcal{G}^2 = 1$.
The triple $(\eta, \mathcal{K}, \omega_{\mathcal{K}})$ where
$\omega_{\mathcal{K}} = \eta \mathcal{K}$, defines a Born structure on
$T \mathcal{M}_{2D}$ \cite{Freidel:2013zga, Freidel:2017yuv, Rudolph:2019fir}.
Given $\mathcal{K}$ and $\mathcal{G}$ that anti-commute with each other
$\{\mathcal{G}, \mathcal{K} \} = 0$ we define $\mathcal{I} = \mathcal{G}
\mathcal{K}$ whose square is minus the identity $\mathcal{I}^2 = - 1$.
In the standard parametrization in type II theories, they are given by
\begin{align}
\mathcal{I} = 
\begin{pmatrix}
0 & - g^{-1} \\
g & 0
\end{pmatrix}
,
\qquad
\mathcal{G}
=
\begin{pmatrix}
0 & g^{-1} \\
g & 0
\end{pmatrix}
,
\qquad
\mathcal{K}
=
\begin{pmatrix}
\mathbf{1}_D & 0 \\
0 & - \mathbf{1}_D
\end{pmatrix}
.
\end{align}
These structures satisfy the algebra of 
a hypercomplex number, the split-quaternion, which we denote
$\mathrm{Sp}\mathbb{H}$ \cite{Kimura:2022jyp}:
\begin{align}
&
- \mathcal{I}^2 
= 
\mathcal{G}^2 
= 
\mathcal{K}^2 
= 1
\qquad 
\mathcal{I} \mathcal{G} \mathcal{K} = 1,
\notag \\
&
\{
\mathcal{I}, \mathcal{G}
\}
=
\{
\mathcal{G}, \mathcal{K}
\}
=
\{
\mathcal{K}, \mathcal{I}
\}
=
0.
\label{eq:Born_algebra}
\end{align}
These relations together with compatibility conditions with $\eta$ and $\mathcal{H}$, the structures $(\mathcal{I},
\mathcal{G}, \mathcal{K})$ characterize the Born manifold.
Note that the introduction of the $B$-field does not change the algebra
since it is given by a similarity transformation.

Under the strong constraint in DFT, the generalized tangent bundle
$\mathbb{T} M_D = TM_D \oplus T^*M_D$ is
identified with the tangent bundle of the doubled space $T\mathcal{M}_{2D}$ \cite{Freidel:2017yuv}.
In the following, we identify $\mathbb{T} M_D = TM_D \oplus T^*M_D$ and 
$T \mathcal{M}_{2D}$ assuming that the strong constraint is always
imposed in any quantities.
Then, any generalized structures on $\mathbb{T} M_D$ are identified with endomorphisms on $T
\mathcal{M}_{2D}$ \cite{Kimura:2022dma, Kimura:2022jyp}.

When the spacetime geometry $M_D$ admits a K\"ahler structure
characterized by $(J,\omega, g)$, this
is promoted to the generalized K\"ahler structure $(\mathcal{J}_J,
\mathcal{J}_{\omega}, \mathcal{G})$ in the doubled space \cite{Gualtieri:2004}.
This satisfies the algebra of bi-complex numbers $\mathbb{C}_2$:
\begin{align}
&
- \mathcal{J}_J^2 = - \mathcal{J}_{\omega}^2 = \mathcal{G}^2 = 1
\qquad
\mathcal{J}_J \mathcal{J}_{\omega} \mathcal{G} = 1,
\notag \\
&
[\mathcal{J}_J, \mathcal{J}_{\omega}]
=
[\mathcal{J}_{\omega}, \mathcal{G}]
=
[\mathcal{G}, \mathcal{J}_J]
=
0.
\label{eq:C2_algebra}
\end{align}
Since $\mathcal{G}$ in the generalized complex structure is a part of
the Born structure, it is shared by both structures.
It has been shown that the structures $\mathcal{J}_J,\mathcal{J}_{\omega}, \mathcal{I},
\mathcal{K}$ and $\mathcal{G}$ do not satisfy a closed algebra, but 
they together with their products satisfy the algebra of the
eight-dimensional bi-quaternions \cite{Kimura:2022jyp}.
In this sense, the Born and the generalized K\"ahler structures are
realized as parts of higher dimensional hypercomplex structures in the doubled space.
The same is true for the hyperk\"ahler structures.

Now we examine the gauge-dressed geometry in heterotic theories.
As discussed in Section~\ref{sect:quasi-compl_and_het_TD}, the effect of the non-Abelian gauge fields in
heterotic supergravity can be incorporated into the $O(D,D)$ 
formalism by the gauge-dressed metric $\bar{g}_{\mu\nu} = g_{\mu \nu} +
\frac{1}{2} \mathrm{Tr} [A_{\mu} A_{\nu}]$.
Given the quasi complex structure $\bar{J}^2 < 0$ satisfying the
relation $\omega = - \bar{g} \bar{J}$, one can naturally introduce an almost
complex structure $\tilde{J} = \frac{1}{\sqrt{\bar{J} \bar{J}^*}}
\bar{J}$ that satisfies $\tilde{J}^2 = -1$ \cite{daSilva:2008}
(see Appendix \ref{App:compatible_comp_str}).
We call this the gauge-dressed almost complex structure.
A compatible metric $\tilde{g}$ is introduced via $\omega = - \tilde{g}
\tilde{J}$ which results in $\omega (\tilde{J} \cdot, \tilde{J} \cdot) = \omega
(\cdot, \cdot)$.
Explicitly, we have $\tilde{g} = \omega \tilde{J} = - \bar{g} \bar{J}
\tilde{J} = \bar{g} \sqrt{\bar{J} \bar{J}^*}$.
Since the metric $\tilde{g}$ is positive-definite, we define a $2D \times 2D$
generalized metric $\widetilde{\mathcal{H}} = \text{diag}
(\tilde{g}, \tilde{g}^{-1})$, which allows us to
analyze the algebraic structures in the doubled space.
It is natural to define the heterotic Born structure in the doubled space by the
triple $(\widetilde{\mathcal{I}}, \widetilde{\mathcal{G}}, \mathcal{K})$
satisfying compatibility conditions with the metrics $\eta$ and
$\widetilde{\mathcal{H}}$.
The para-complex structure $\mathcal{K}$ is defined as in the type II
case while the chiral structure $\widetilde{\mathcal{G}}$ and
$\widetilde{\mathcal{I}}$ are defined through $\widetilde{\mathcal{H}}$ as
\begin{align}
\widetilde{\mathcal{I}} 
= 
\widetilde{\mathcal{G}}
\mathcal{K}
=
\begin{pmatrix}
0 & - \tilde{g}^{-1} \\
\tilde{g} & 0
\end{pmatrix}
,
\quad
\widetilde{\mathcal{G}} 
= 
\eta^{-1} 
\widetilde{\mathcal{H}} 
= 
\begin{pmatrix} 
0 & \tilde{g}^{-1} 
\\ 
\tilde{g} & 0 
\end{pmatrix}, 
\quad
\mathcal{K} = 
\begin{pmatrix} 
1 & 0 \\ 
0 & -1 
\end{pmatrix},
\end{align}
where we have employed the canonical basis $\mathbb{X}^M = (x^{\mu},
\tilde{x}_{\mu})$.
It is obvious that the structures satisfy the $\mathrm{Sp}\mathbb{H}$ algebra \eqref{eq:Born_algebra}.
The $B$-field is introduced into these structures through the $B$-transformation such as 
$\widetilde{\mathcal{H}}{}^B = (\e^B)^{\Transpose} \widetilde{\mathcal{H}} \e^B$, where $\e^B$ is
defined in \eqref{eq:B-transf_mat}.
Given a gauge-dressed almost complex structure $\tilde{J}$, 
we define the gauge-dressed generalized (almost) complex structures in the doubled
space as
\begin{align}
\widetilde{\mathcal{J}}_{J} 
= 
\begin{pmatrix} 
\tilde{J} & 0 \\ 
0 & -\tilde{J}^* 
\end{pmatrix}, 
\qquad
\mathcal{J}_{\omega} = 
\begin{pmatrix} 
0 & -\omega^{-1} 
\\ 
\omega & 0 
\end{pmatrix}
.
\end{align}
Note that we have $\widetilde{\mathcal{J}}_J^2 = - 1$ but $\overline{\mathcal{J}}_{\bar{J}}^2 \not= -1$.
Note also that  
$\mathcal{J}_{\omega}$ is integrable while
$\widetilde{\mathcal{J}}_J$ is generically not, hence the prefix ``almost.''
It is obvious that the triple 
$(\widetilde{\mathcal{J}}_{J}, \mathcal{J}_{\omega}, \widetilde{\mathcal{G}})$ 
 satisfies the algebra of bi-complex numbers $\mathbb{C}_2$, \eqref{eq:C2_algebra}.
It is also easy to show that $\tilde{\mathcal{J}}_{J}$ and $\mathcal{J}_{\omega}$ commute
with each other and $\widetilde{\mathcal{G}}$ is positive-definite.
Then they indeed define a generalized almost K\"ahler structure
\cite{Gualtieri:2004}.

We next consider the embedding of the $(p,q)$-hermitian
structure in the heterotic Born geometry.
The spacetime $M_D$ admits $(p-1)$ gauged-dressed almost complex structures $\tilde{J}_{+,a}$
($a=1, \dots, p-1$) and also $(q-1)$ $\tilde{J}_{-,a'}$ ($a'=1,
\dots, q-1$).
As we have clarified, they are introduced via the gauge-dressed metric
$\bar{g}$ and the symplectic 2-forms $\omega_{+,a}, \omega_{-,a'}$.
We stress that even if the original $J_{\pm,a}$ satisfy the
hyperk\"ahler relation $J_{\pm,1} J_{\pm,2} = J_{\pm,3}$, the
gauge-dressed almost complex structure $\tilde{J}_{\pm,a}$ does not
satisfy it in general.

We are interested in the algebra that accommodates 
the heterotic Born and the generalized K\"ahler structures
associated with the $(p,q)$-hermitian geometry.
The cases $(p,q) = (4,4), (2,2)$ and $(4,0), (2,0)$ correspond to the
bi-hypercomplex, bi-hermitian, hyperk\"ahler, and K\"ahler geometries,
respectively and their doubled counterparts in type~II theories are discussed in~\cite{Kimura:2022jyp}.
We here focus on the $(p,q) = (4,2)$ case which is a non-trivial example in 
heterotic theories and corresponds to a mixing of the geometry with the
hyperk\"ahler and K\"ahler structures.
The left sector ($p=4$) admits a hyperk\"ahler structure consisting of
three complex structures $J_{+,a}$ ($a=1,2,3$).
Given $J_{+,a}$ and $\omega_{+,a}$, we can construct $\tilde{J}_{+,a}$ and
$\tilde{g}_{+,a}$ and they are embedded into the 
three independent $\mathbb{C}_2$ structures
$(\widetilde{\mathcal{J}}_{J_+,a},
\mathcal{J}_{\omega_+,a}, 
\widetilde{\mathcal{G}}_{+,a})$ ($a=1,2,3$).
We again stress that $\widetilde{\mathcal{J}}_{J_+,a}$ and
$\mathcal{J}_{\omega_+,a}$ do not obey the algebra~\eqref{eq:generalized_hyperkahler_algebra} in general.
Furthermore, we have independent metrics $\tilde{g}_{+,a}$ for each $\tilde{J}_{+,a}$.
The right sector ($q=2$) admits a single gauge-dressed almost complex structure
$\tilde{J}_{-}$ and $\tilde{g}_{-}$, which form another $\mathbb{C}_2$ structure
$
(\widetilde{\mathcal{J}}_{J_-}, 
\mathcal{J}_{\omega_-},
\widetilde{\mathcal{G}}_{-})
$.
See Table~\ref{tb:structures}.

\begin{table}[t]
\centering
\renewcommand{\arraystretch}{1.5}
\begin{tabular}{|c|c|c|c|}
\hline
\textbf{type} & \textbf{spacetime} & \textbf{doubled space} & \textbf{algebra} 
\\ 
\hline
heterotic
& 
& Born $(\widetilde{\mathcal{I}}, \widetilde{\mathcal{G}}, \mathcal{K} )$
& split-quaternions $\mathrm{Sp} \mathbb{H}$
\\
\cline{2-4}
& hyperk\"ahler 
& GACS $(\tilde{\mathcal{J}}_{J_+,a}, \mathcal{J}_{\omega_+,a}, \widetilde{\mathcal{G}}_{+,a})$ 
& bi-complex numbers $\mathbb{C}_2$ 
\\ 
\cline{2-4}
& K\"ahler
& GACS $(\tilde{\mathcal{J}}_{J_-}, \mathcal{J}_{\omega_-}, \widetilde{\mathcal{G}}_{-})$ 
& bi-complex numbers $\mathbb{C}_2$ 
\\ 
\hline
\hline
type II
& 
& Born $(\mathcal{I}, \mathcal{G}, \mathcal{K})$
& split-quaternions $\mathrm{Sp} \mathbb{H}$
\\
\cline{2-4}
& hyperk\"ahler 
& GHKS $(\mathcal{J}_{J,a}, \mathcal{J}_{\omega,a}, \mathcal{G})$ 
& split-bi-quaternions $\mathrm{Sp}\mathbb{C} \times \mathbb{H}$
\\ \cline{2-4}
& K\"ahler
& GCS $(\mathcal{J}_{J}, \mathcal{J}_{\omega}, \mathcal{G})$ 
& bi-complex numbers $\mathbb{C}_2$ 
\\
\hline
\end{tabular}
\caption{Geometric structures in spacetime and doubled space and
 associated algebras.
The heterotic and type II theories are shown for comparison.
Each abbreviation stands for GACS (generalized almost complex
 structure), GHKS (generalized hyperk\"ahler structure), GCS
 (generalized complex structure), respectively.
}
\label{tb:structures}
\end{table}

We expect a geometric structure that involves the Born and the four generalized
almost complex structures. 
Indeed, such a geometric structure exists in the type~II case~\cite{Kimura:2022dma}. 
In that case, the Born and the generalized (hyper)K\"ahler structures
share a common chiral structure $\mathcal{G}$. 
However, in the heterotic case, distinct chiral structures
$\widetilde{\mathcal{G}}_{+,a}, \widetilde{\mathcal{G}}_-$ exist for
each gauge-dressed $\tilde{J}_{+,a}, \tilde{J}_-$, and consequently, the
algebraic structures associated with each structure are completely
independent.
Therefore, heterotic theories do not admit such 
a geometric structure that accommodates the Born and the generalized almost
complex structures as is found in Type~II.

\section{An example of gauge-dressed geometry} \label{sec:example}
In this section, we show an explicit example of gauge-dressed geometry.
The so-called symmetric five-brane background in heterotic supergravity
is given by \cite{Callan:1991dj, Callan:1991ky}
\begin{align}
{\dop}s^2 &=  {\dop}s_6^2 + H (r) \, {\dop}s_4^2,
\notag \\
\e^{2\phi} &= H (r) = \e^{2 \phi_0} + \frac{n \alpha'}{r^2},
\quad
3 \del_{[\mu} B_{\nu \rho]} = \varepsilon_{\mu \nu \rho \sigma}
 \del^{\sigma} H,
\notag \\
A_{\mu} &= - {\I} \frac{\sigma_{\mu \nu} x^{\nu}}{r^2 + n \alpha' \e^{-2
 \phi_0}},
\label{eq:5-brane}
\end{align}
where ${\dop}s^2_6$, ${\dop}s^2_4$ are the six-dimensional worldvolume and
transverse directions of the five-branes.
$x^{\mu}$ ($\mu = 1,2,3,4$), $r^2 = \delta_{\mu \nu} x^{\mu} x^{\nu}$
are coordinates in the transverse directions, $n,\phi_0$ are constants.
In the following, we focus on the four-dimensional transverse space.
$H$ is a harmonic function in the transverse space.
The solution preserves $\mathcal{N} = (4,4)$ supersymmetry and hence it
is an example of the $(4,4)$-hermitian geometry and 
it admits a bi-hypercomplex structure.
The gauge field satisfying the self-duality condition in the four
dimensions takes its value in the SU(2) subsector of the gauge
group SO(32) or $\mathrm{E}_8 \times \mathrm{E}_8$.
This four-dimensional space admits a bi-hypercomplex structure and the
complex structures are given by
\begin{align}
    J_{1,+} &= \begin{pmatrix} 0 & {\I}\sigma_2 \\ {\I}\sigma_2 & 0 \end{pmatrix}, & \hspace{-20pt}
    J_{2,+} &= \begin{pmatrix} 0 & \mathbf{1}_2 \\ -\mathbf{1}_2 & 0 \end{pmatrix}, & \hspace{-20pt}
    J_{3,+} &= \begin{pmatrix} -{\I}\sigma_2 & 0 \\ 0 & {\I}\sigma_2 \end{pmatrix}, \notag \\
    J_{1,-} &= \begin{pmatrix} 0 & -\sigma_1 \\ \sigma_1 & 0 \end{pmatrix}, & \hspace{-20pt}
    J_{2,-} &= \begin{pmatrix} 0 & \sigma_3 \\ -\sigma_3 & 0 \end{pmatrix}, & \hspace{-20pt}
    J_{3,-} &= \begin{pmatrix} -{\I}\sigma_2 & 0 \\ 0 & -{\I}\sigma_2 \end{pmatrix},
\label{eq:5-brane_J}
\end{align}
which is the same as that in the NS5-brane geometry in type II
theories \cite{Papadopoulos:2000iv}.
It is easy to confirm that these satisfy the hermitian condition with respect to the
four-dimensional metric $g_4$.
It is known that a T-duality transformation of the solution
\eqref{eq:5-brane} gives another five-brane \cite{Sasaki:2016hpp,
Sasaki:2017yrs}.
For this geometry, the gauge-dressed metric is defined by
\begin{align}
\bar{g}_{\mu \nu} =& \ g_{\mu \nu} + \frac{1}{2} \mathrm{Tr} A_{\mu}
 A_{\nu}
\notag \\
=& \ H \delta_{\mu \nu} + \frac{1}{16 f^2} 
\Big(
\delta_{\mu \nu} \del_{\rho} f \del_{\rho} f
+
\del_{\mu} f \del_{\nu} f
\Big),
\label{eq:gauge-shefted_NS5}
\end{align}
where $f = r^2 + n \alpha' \e^{-2 \phi_0}$ is the function specifying the
self-dual solution $A_{\mu} = - \frac{{\I}}{2} \sigma_{\mu \nu} \del_{\nu} \log (c f)$ 
($c$ is a dimensionful constant).
We assume that $\bar{g}$ is non-degenerate.
The symplectic 2-forms are given by
$(\omega_{a,\pm})_{\mu \nu} = - g_{\mu \rho} (J_{a,\pm})^{\rho}{}_{\nu}$
($a = 1,2,3$).
Given a symplectic form $\omega$, and the Riemannian metric $\bar{g}$, we
can construct a quasi complex structure $\bar{J}^{\mu} {}_{\nu}$ through
the relation $\bar{J}_{a,\pm} = - \bar{g}^{-1} \omega_{a,\pm}$.
The gauge-shifted metric in \eqref{eq:gauge-shefted_NS5} is rewritten as 
\begin{align}
\bar{g}_{\mu \nu}
= \frac{1}{16 f^2}
\left\{
(\lambda + s) \delta_{\mu \nu} + u_{\mu} u_{\nu}
\right\},
\end{align}
where we have defined $u_{\mu} = \del_{\mu} f$, $s = u^2 = \delta^{\mu \nu}
u_{\mu} u_{\nu}$.
Its inverse is found to be
\begin{align}
\bar{g}^{\mu \nu} = \frac{16 f^2}{\lambda + s} 
\left\{
\delta^{\mu \nu} - \frac{u^{\mu} u^{\nu}}{\lambda + 2 s}
\right\}.
\end{align}
Here the indices are raised and lowered by $\delta^{\mu \nu}$ and
$\delta_{\mu \nu}$.
Then we find that the quasi complex structures are given by
\begin{align}
(\bar{J}_{a,\pm})^{\mu} {}_{\nu} =& \ 
- \bar{g}^{\mu \rho} (\omega_{a,\pm})_{\rho \nu}
\notag \\
=& \ \bar{g}^{\mu \rho} g_{\rho \kappa} (J_{a,\pm})^{\kappa} {}_{\nu}
\notag \\
=& \ 
\frac{\lambda}{\lambda + s}
\left\{
(J_{a,\pm})^{\mu} {}_{\nu}
- \frac{1}{\lambda + 2 s} u^{\mu} u_{\rho} (J_{a,\pm})^{\rho} {}_{\nu}
\right\}.
\end{align}
It is obvious that when $A_{\mu} = 0$ ($u_{\mu} = \del_{\mu} f = 0$), the quasi
complex structures $\bar{J}_{a,\pm}$ reduce to the complex structures $J_{a,\pm}$.
Using $\bar{J}_{a,\pm}$, we can construct
gauge-dressed almost complex structures $\tilde{J}_{a,\pm}$:
\begin{align}
\tilde{J}_{a,\pm} = \Big( \sqrt{\bar{J}_{a,\pm} \bar{J}^{*}_{a,\pm}}
 \Big)^{-1} \bar{J}_{a,\pm} \qquad (\text{no sum over $a$}), 
\label{eq:gdJ_NS5}
\end{align}
where $\bar{J}^*_{a,\pm}$ is the adjoint of $\bar{J}_{a,\pm}$ with
respect to $\bar{g}$, namely, $\bar{g} (\bar{J}_{a,\pm} \cdot, \cdot) = 
\bar{g} (\cdot, \bar{J}^*_{a,\pm} \cdot)$ or in the matrix notation $\bar{J}^*_{a,\pm} =
\bar{g}^{-1} \bar{J}^{\Transpose}_{a,\pm} \bar{g}$.
Since we have $\bar{J}^* = \bar{g}^{-1} \omega \bar{g}^{-1} \bar{g} =
\bar{g}^{-1} \omega = - \bar{J}$, we find $\bar{J} \bar{J}^* = -
\bar{J}^2 $ for each $J_{a,\pm}$.
This is explicitly calculated as
\begin{align}
(\bar{J} \bar{J}^{*})^{\mu} {}_{\nu}
=
- (\bar{J}^2)^{\mu} {}_{\nu}
=
\left(
\frac{\lambda}{\lambda + s}
\right)^2
\left[
\delta^{\mu} {}_{\nu}
-
\frac{u^{\mu} u_{\nu} - v^{\mu} v^t_{\nu}}{\lambda + 2 s}
\right]
\qquad
\text{for each $J_{a,\pm}$}.
\end{align}
Here we have used the fact $u_{\alpha} J^{\alpha} {}_{\rho} u^{\rho} =
0$ by the skew-symmetric nature of $\omega_{\mu \nu} = - g_{\mu \rho}
J^{\rho} {}_{\sigma} = - H \delta_{\mu \rho} J^{\rho} {}_{\sigma} = - H
J_{\mu \sigma}$ and defined $s = u^2$, $v^{\mu} = J^{\mu} {}_{\rho} u^{\rho}$,
$v^t_{\mu} = u_{\beta} J^{\beta} {}_{\mu}$.
Noticing the orthogonality of $u_{\mu}$ and $v^{\mu}$, we find the
explicit form of the square root of the matrix $\bar{J} \bar{J}^*$ and its inverse:
\begin{align}
\left(
\sqrt{\bar{J} \bar{J}^{*}}
\right)^{\mu} {}_{\nu}
=& \ 
\frac{\lambda}{\lambda + s}
\left[
\delta^{\mu} {}_{\nu} + 
\left(
\sqrt{\frac{\lambda + s}{\lambda + 2 s}} - 1
\right)
\frac{u^{\mu} u_{\nu} - v^{\mu} v^t_{\nu}}{s}
\right],
\notag \\
\Big(
\sqrt{\bar{J} \bar{J}^{*}}^{-1}
\Big)^{\mu} {}_{\nu}
=& \
\frac{\lambda + s}{\lambda}
\left[
\delta^{\mu} {}_{\nu} 
+
\left(
\sqrt{\frac{\lambda + 2 s}{\lambda + s}} - 1
\right)
\frac{u^{\mu} u_{\nu} - v^{\mu} v^t_{\nu}}{s}
\right]
\qquad
\text{for each } J_{a,\pm},
\end{align}
and then \eqref{eq:gdJ_NS5} is obtained.

By construction, $\tilde{J}_{a,\pm}$ is orthogonal $\tilde{J}_{a,\pm}
\tilde{J}^*_{a,\pm} = 1$ and $\tilde{J}^2_{a,\pm} = -1$. 
We emphasize that the fact that $J_{a,\pm}$ is integrable does not imply that
$\tilde{J}_{a,\pm}$ is integrable.
This is confirmed by evaluating the Nijenhuis tensor
$
N_{\tilde{J}} (\cdot, \cdot) = - \tilde{J}^2 [\cdot, \cdot] + 
\tilde{J} 
\Big(
[\tilde{J} \cdot, \cdot]
+
[\cdot \tilde{J} \cdot]
\Big)
-
[\tilde{J} \cdot, \tilde{J} \cdot]
$
associated with $\tilde{J}$ and clarifying whether it is zero or not.

We note that the symmetric five-brane background~\eqref{eq:5-brane} actually holds under the standard embedding ansatz~\cite{Candelas:1985en}. 
This is the condition that identifies the gauge field $A_{\mu}$ with the Lorentz spin connection $\omega_{\mu}$. 
Under this condition, the shift by the gauge field in $\bar{g}$ is canceled by the 
higher-derivative corrections stemming from $\omega_{\mu}$. 
We will mention this issue in the next section.

\section{Conclusion and discussions} \label{sec:conclusion}

In this paper, we have investigated the T-duality transformations of
$(p,q)$-hermitian geometries in the presence of non-Abelian gauge fields
in heterotic string theories. 
In the absence of gauge fields, the $(p,q)$-hermitian 
geometries appear as target spaces of two-dimensional $\mathcal{N}=(p,q)$ supersymmetric
sigma models and the associated generalized geometries have been
studied.
We focused on the interplay between the generalized geometry and
the doubled formalism and we utilized them to show the $\group{O}(D,D)$ T-duality
transformation rules for geometric quantities.
When there are non-zero gauge field backgrounds, the T-duality transformation generically
receives corrections from them.
We introduced a notion of the gauge-dressed geometry
and showed that a natural basis for T-duality in heterotic theories is not the original spacetime
metric $g_{\mu\nu}$, but a shifted metric $\bar{g}_{\mu\nu} =
g_{\mu\nu} + \frac{1}{2} \mathrm{Tr} (A_{\mu} A_{\nu})$.
Under this shift, we defined the
quasi complex structures $\bar{J}_{\pm}$ defined by
the fundamental two-forms $\omega_{\pm} = - \bar{g} \bar{J}_{\pm}$ through the
structures $(g,J_{\pm},\omega_{\pm})$ of the original complex geometry.
This construction ensures that the compatibility between the metric and
the complex structures is preserved in a way that is consistent with the
$\group{O}(D,D)$ transformation rules.
Utilizing the doubled formalism and the $\group{O}(D,D+n)$ to $\group{O}(D,D)$
reduction, we derived the explicit heterotic Buscher-like rules
for the geometric quantities in heterotic theories. 
We found that the complex structures
$(J_{+,a}, J_{-,a'})$ transform non-trivially, receiving shifts
from the non-Abelian gauge fields. 
The result is consistent with those in the literature 
\cite{Hassan:1994mq} and generalizes the previous results known in Type II theories.

We then introduced the gauge-dressed almost complex structure $\tilde{J}$
through the quasi complex structure $\bar{J}$ and the fundamental 2-form
$\omega$. 
This does satisfy $\tilde{J}^2 = -1$ and we showed that the gauge-dressed structures are naturally
embedded into a Born geometry in the doubled space.
We analyzed the algebraic structures of the doubled space and found that the
$(p,q)$-generalized complex structures, together with the heterotic Born structure
$(\widetilde{\mathcal{I}}, \widetilde{\mathcal{G}}, \mathcal{K})$, satisfy
hypercomplex algebras. 
Specifically, for the $(p,q) = (4,2)$ case, the geometry has hypercomplex
structures satisfying the algebra of $\mathbb{C}_2$ which is different from that in type II
case.
We also exhibited an example of the gauge-dressed geometry and
wrote down the explicit quasi and gauge-dressed almost complex structures.

There are several directions for future research. 
While we focused on the gauge sector in the 
$\mathcal{O}(\alpha')$ corrections, the
inclusion of the local Lorentz sector is also important since it gives
the Riemann square term $R^2$, which is a higher derivative correction including fourth derivatives in the heterotic supergravity action.
The anomaly cancellation suggests that the spacetime metric $g_{\mu
\nu}$ is replaced by $\bar{g}_{\mu \nu} = g_{\mu \nu} + \frac{1}{2} \mathrm{Tr}
[ A_{\mu} A_{\nu}] - \frac{1}{2} \mathrm{Tr} [\omega_{\mu} \omega_{\nu}]$ where $\omega_{\mu}$ is
the spin connection for the local Lorentz group~\cite{Hull:1986xn,Sen:1986nm}.
Incorporating the spin connection as a 
gauge field of the local Lorentz group would be a
natural extension 
of the gauge-shifted metric discussed in this paper~\cite{Bedoya:2014pma}.
It would also be interesting to apply our
Buscher-like rules to explicit non-geometric backgrounds in heterotic
theories~\cite{Sasaki:2016hpp, Sasaki:2017yrs, Kimura:2023nvt}.
The applications of the heterotic Born sigma model to the integrable
deformations are also interesting~\cite{Osten:2023cza}.
Finally, the connection between the $(p,q)$-generalized geometry and
higher-dimensional duality covariant field theories~\cite{Malek:2017njj,
Hassler:2023nht, Hassler:2024yis} offers a promising path toward a unified understanding
of string dualities.
It would be also interesting to exhibit explicit solutions that have
non-trivial complex structures by using the T-duality transformation in
heterotic string theories.
We will come back these issues in future studies.

\subsection*{Acknowledgments}
The authors would like to thank Haruka Mori for her valuable
contributions in the early stage of this work.
The work of S.~S.\ and K.~S.\ is supported by Grant-in-Aid for Scientific
Research, JSPS KAKENHI Grant Number JP25K07324. 
The work of K.~S.\ is also supported by Kitasato University Research Grant for Young Researchers. 


\begin{appendix}

\appendix

\section{Compatible almost complex structures}
\label{App:compatible_comp_str}
In this appendix, we briefly introduce the construction of a compatible
complex structure $\tilde{J}$ from the quasi complex structure $\bar{J}$.
Details and a rigorous proof can be found, for example, in~\cite{daSilva:2008}.

Let $(V,\omega)$ be a finite-dimensional symplectic vector space,
{i.e.}, $\omega$ is a non-degenerate, skew-symmetric bilinear map.
We choose an arbitrary positive-definite inner product (metric) $\bar{g}$
on $V$. 
Since both $\bar{g}$ and $\omega$ are non-degenerate, they induce isomorphisms between $V$ and $V^*$.
Therefore, there exists a unique linear map $\bar{J} : V \to V$ such that
\begin{equation}
    \omega(u,v) = \bar{g}(\bar{J} u,v)
    \label{eq:A-definition}
\end{equation}
for all $u,v\in V$.
In matrix notation, this is written as $\omega = \bar{J}^{\Transpose} \bar{g} = -
\bar{g} \bar{J}$,
and hence $\bar{J} = - \bar{g}^{-1} \omega$ where we used $\bar{g}^{\Transpose} =
\bar{g}$ and $\omega^{\Transpose} = -\omega$.
Then we have the relation,
\begin{align}
    \bar{g}(\bar{J}^*u,v)
    &= \bar{g}(u,\bar{J}v) \notag \\
    &= \bar{g}(\bar{J}v,u) \notag \\
    &= -\omega(u,v) \notag \\
    &= \bar{g}(-\bar{J}u,v), \quad \text{for } {}^{\forall} u,
 {}^{\forall} v,
\end{align}
where $\bar{J}^*$ is the adjoint of $\bar{J}$ with respect to $\bar{g}$. 
Then we have $\bar{J}^* = - \bar{J}$.

Now consider the operator $\bar{J}\bar{J}^*$.
It is self-adjoint, since $(\bar{J} \bar{J}^*)^*= \bar{J} \bar{J}^*$
and is positive definite,
\begin{equation}
    \bar{g}(\bar{J} \bar{J}^*u,u) = \bar{J}(\bar{J}^*u,\bar{J}^*u)>0
    \qquad
    (u\neq 0).
\end{equation}
Here we have used the positive definiteness of $\bar{g}$ and the non-degeneracy of $\bar{J}$.
Since $\bar{J}\bar{J}^* > 0$, it admits a positive square root.
We also find $\bar{J}^2 = - \bar{J} \bar{J}^* <0$.

Now we define
\begin{equation}
    \tilde{J} = \left(\sqrt{\bar{J}\bar{J}^*}\right)^{-1} \bar{J},
    \label{eq:J-polar}
\end{equation}
or equivalently,
\begin{equation}
    \bar{J}=\sqrt{\bar{J}\bar{J}^*}\,\tilde{J},
\end{equation}
which is called the polar decomposition of $\bar{J}$.
Then we find
\begin{align}
    \tilde{J}\tilde{J}^*
    &=
    \left(\sqrt{\bar{J} \bar{J}^*}\right)^{-1}
    \bar{J} \bar{J}^*
    \left(\sqrt{\bar{J}\bar{J}^*}\right)^{-1}
    \notag \\
    &=
    (\bar{J}\bar{J}^*)^{-1}\bar{J}\bar{J}^*
    \notag \\
    &=1,
\label{eq:app1}
\end{align}
where we have used the fact that $\bar{J}$ commutes with $\sqrt{\bar{J} \bar{J}^*}$.
We also find that 
\begin{align}
    \bar{g}(\tilde{J}^*u,v)
    &= \bar{g} (\tilde{J}v,u) \notag \\
    &= \bar{g} \big( \sqrt{\bar{J}\bar{J}^*}^{-1}
 \bar{J} v,u \big) \notag \\
    &= \omega \big( \sqrt{\bar{J} \bar{J}^*}^{-1} v , u \Big) \notag \\
    &= - \bar{g} \big(\bar{J} v, \sqrt{\bar{J}\bar{J}^*}^{-1} u \big) \notag \\
    &= -\bar{g} \big(v,\bar{J}^* \sqrt{\bar{J}\bar{J}^*}^{-1}u \big) \notag \\
    &= -\bar{g} (v,\bar{J}^* u) \notag \\
    &= \bar{g}(-\bar{J} u,v), \quad \text{for } {}^{\forall} u, {}^{\forall} v.
\end{align}
Therefore we have $J^*=-J$ and combining this with \eqref{eq:app1}, we conclude
\begin{equation}
    \tilde{J}^2 = - \tilde{J} \tilde{J}^* = -1.
\end{equation}
We can verify that $\tilde{J}$ is compatible with $\omega$. 
Indeed, we have
\begin{align}
    \omega(\tilde{J} u, \tilde{J} v)
    &= \bar{g}(\bar{J} \tilde{J} u, \tilde{J}v) \notag \\
    &= \bar{g}(\tilde{J} \bar{J} u, \tilde{J}v) \notag \\
    &= \bar{g}(\bar{J}u,\tilde{J}^* \tilde{J}v) \notag \\
    &= \bar{g}(\bar{J}u,v) \notag \\
    &= \omega(u,v),
\end{align}
where we have again used the fact that $\bar{J}$ and $\tilde{J}$ commute
with each other and $\tilde{J}^* \tilde{J}=1$.
Next, 
\begin{align}
    \omega(u,\tilde{J} u)
    &= \bar{g} (\bar{J} u, \tilde{J} u) \notag \\
    &= \bar{g}(\tilde{J}^* \bar{j} u,u) \notag \\
    &= \bar{g}(-\tilde{J} \bar{J} u,u) \notag \\
    &= \bar{g} \left(-\bar{J} \sqrt{\bar{J}\bar{J}^*} \tilde{J}u,u\right) \notag \\
    &= \bar{g} \left( \sqrt{\bar{J} \bar{J}^*}u,u\right)>0
    \qquad
    (u\neq 0).
\end{align}
Thus $\tilde{J}$ is $\omega$-compatible.
It is easy to generalize this construction for a Riemannian metric
$\bar{g}$ and a symplectic 2-form $\omega$ on a manifold $M$.
Therefore $\tilde{J}$ is an almost complex structure on $M$.
Note that the positive-definite metric $\tilde{g} (u,v) = \omega (u, \tilde{J}v)$
is in general not equal to the initially chosen metric $\bar{g}$.

\end{appendix}


\end{document}